\documentclass{article}
\usepackage[preprint]{template}

\usepackage[utf8]{inputenc} 
\usepackage[T1]{fontenc}    
\usepackage{hyperref}       
\usepackage{url}            
\usepackage{booktabs}       
\usepackage{amsmath}        
\usepackage{amsthm}         
\usepackage{amsfonts}       

\usepackage{nicefrac}       
\usepackage{microtype}      
\usepackage{xcolor}         
\usepackage{graphicx}
\usepackage{multirow}
\usepackage{array}
\title{Policy-DRIFT: Dynamic Reward-Informed Flow Trajectory Steering}

\author{%
  Atharva Mahajan\thanks{Corresponding authors: \texttt{atharvm@umich.edu}, \texttt{abvish@umich.edu}, \texttt{rvinuesa@umich.edu} \\ $^\dagger$ Equal contribution.}$^{*\dagger}$ \\
  Department of Aerospace Engineering\\
  University of Michigan, Ann Arbor \\
  \texttt{atharvm@umich.edu} \\
  \And
  Abhijeet Vishwasrao$^{*\dagger}$ \\
  Department of Aerospace Engineering\\
  University of Michigan, Ann Arbor \\
  \texttt{abvish@umich.edu} \\
  \And
  Yuning Wang \\
  Department of Aerospace Engineering\\
  University of Michigan, Ann Arbor \\
  \texttt{yuninw@umich.edu} \\
  \And
  Ricardo Vinuesa$^*$ \\
  Department of Aerospace Engineering\\
  University of Michigan, Ann Arbor \\
  \texttt{rvinuesa@umich.edu} \\
}

\begin{document}

\maketitle

\begin{abstract}
Skin-friction drag induced by wall-bounded turbulent flows accounts for a substantial fraction of energy consumption across commercial aerospace, wind energy, and marine transport. Its active reduction is one of the highest-value targets in engineering fluid dynamics. Deep reinforcement learning (DRL) has emerged as the leading approach for real-time flow control, yet its performance ceiling is set not by algorithmic capability but by reward structure, the naive scalar objective does not optimally reflect the underlying physics. Policy-DRIFT bypasses this ceiling by relocating reward information from policy gradients to generative model inference: a conditional flow matching model (CFM) constructs a physically-grounded manifold of realisable flow states spanning multiple control regimes, Terminal Reward Guidance (TRG) steers samples toward reward-maximising targets at inference, and a lightweight DRL policy, structurally decoupled from reward quality, tracks these full-field targets via root-mean-squared error (RMSE) minimisation. The test case is turbulent channel flow simulated using direct numerical simulation (DNS) at friction Reynolds number of $\mathrm{Re}_\tau = 180$, which is the canonical benchmark for wall-bounded turbulence. Policy-DRIFT achieves $49\%$ drag reduction approaching the theoretical upper bound, which is $\approx 16\%$ higher than the DRL benchmark, while consuming 37$\times$ less actuation energy. Our approach combines generative methods with active flow control, marking a paradigm shift towards controlling complex physical systems efficiently.
\end{abstract}

\section{Introduction}
Active control of complex physical systems is central to modern engineering: from stabilising plasma in tokamak reactors to regulating flow in pipeline networks and cardiovascular devices, the ability to manipulate dynamical processes in real time translates directly into performance gains and energy savings. Deep reinforcement learning (DRL) has emerged as the leading paradigm for this challenge, delivering state-of-the-art controllers across robotic manipulation \citep{luo_serl_2025}, molecular design \citep{olivecrona_molecular_2017}, plasma stabilisation \citep{degrave_magnetic_2022}, and turbulent flow control \citep{guastoni_deep_2023, beneitez2025improving}. Wall-bounded turbulent flows represent a particularly high-value target: skin-friction drag accounts for roughly half of total aerodynamic drag on commercial aircraft \citep{vinuesa_AIrole_2020}, and even a few percent reduction helps towards improving global fuel consumption efficiencies and $CO_2$ emissions significantly. Recently, DRL controllers have demonstrated sustained drag reduction systematically outperforming classical heuristics such as opposition control \citep{Choi_Moin_Kim_1994}.

Yet a fundamental limitation persists across DRL applications regardless of algorithmic advances, the reward signal remains a proxy of the true objective, and the policy ceiling is set entirely by how well this surrogate reflects what optimal actually looks like \citep{skalse_defining_2022, pan_effects_2021}.
In high-fidelity physical simulation, this is compounded by the prohibitive cost of sustained online DNS interaction during training, restricting policy improvement to what the proxy reward allows \citep{dulac-arnold_challenges_2019}.
Prior remedies through inverse reward design \citep{hadfield-menell_inverse_2020} and reward-hacking mitigation \citep{laidlaw_correlated_2024} refine the proxy without eliminating the underlying problem: as long as reward gradients flow directly through the policy, learned behaviour remains fundamentally coupled to surrogate quality.

\textbf{Policy-DRIFT} addresses reward misspecification by relocating where reward acts: not as a per-step gradient shaping the DRL policy, but as guidance over a generative model of the controlled system.
A conditional flow matching (CFM) model trained across a spectrum of control strategies constructs a physically-grounded manifold of realisable flow states.
Terminal Reward Guidance (TRG) then navigates this manifold at inference, steering ODE trajectories toward terminal states that maximise a designer-specified reward, fully decoupling the policy's learning signal from reward quality.
A lightweight DRL policy learns exclusively to track these generative targets via RMSE minimisation. Its learning signal is a full-field spatial objective, never a spatially averaged reward gradient, so that optimal behaviour emerges from target tracking rather than reward engineering. Crucially, the CFM model requires no retraining when the control objective changes.

Instantiated with a cost-aware objective on turbulent channel flow DNS at a moderate friction Reynolds number of $\mathrm{Re}_\tau = 180$ ($\mathrm{Re}_\tau = u_\tau h / \nu$ where $u_\tau$ is the friction velocity, $h$ is the channel height and $\nu$ is the viscosity.), Policy-DRIFT outperforms all available baselines: it achieves $48.95\%$ drag reduction, being {$16.2\%$} higher than vanilla DRL and {$38.9\%$} higher than opposition control, while consuming approximately ${\sim}37\times$ less actuation energy than DRL.
Unlike standard DRL, which reacts to instantaneous scalar reward signals, Policy-DRIFT operates as a receding-horizon controller: at each control horizon, CFM with TRG computes the physically-realisable, reward-maximising target one horizon ahead; the DRL policy then executes 8 actuation steps to track it.
Note that neither drag reduction nor energy savings are directly optimised by the policy, which emerge implicitly from target tracking.

Our contributions are as follows:

\begin{itemize}

  \item \textbf{Policy-DRIFT framework.}
  We introduce Policy-DRIFT, a generative control framework that replaces naive DRL reward signals with physically-grounded target states via CFM trained jointly on uncontrolled, opposition-controlled, and DRL-controlled flow regimes, enabling flexible reward specification without reward gradients entering the policy.

  \item \textbf{Terminal Reward Guidance (TRG).}
  We propose TRG, which steers CFM ODE trajectories toward reward-maximising terminal states via gradients of a learned reward predictor, extending inference-time guidance to full-field PDE-governed state spaces with a manifold-regularising pre-placement design that prevents reward hacking without constraining the reward objective.

  \item \textbf{Generative target generation for turbulence control.}
  CFM combined with TRG provides physically-grounded, reward-maximising flow targets
  during training, decoupling target discovery from policy execution such
  that the deployed policy acts reactively from wall-parallel sensing alone,
  without generative model queries at inference time.

  \item \textbf{Empirical validation on DNS benchmark.}
  On $\mathrm{Re}_\tau = 180$ turbulent channel flow, Policy-DRIFT simultaneously achieves $16.2\%$ higher drag reduction than standard DRL and $38.9\%$ higher than opposition control, while using approximately ${\sim}37\times$ less actuation energy than DRL, with both gains emerging without the policy directly optimising either quantity.

\end{itemize}

\section{Related Work}

\textbf{Turbulent flow control. }
\citet{bewley2001dns} established that receding-horizon optimal control
with full DNS access achieves over 50\% drag reduction and complete flow
relaminarisation in turbulent channel flow at $\mathrm{Re}_\tau = 180$,
setting the physical upper bound for wall-transpiration control, but at
the cost of iterative full-state DNS at every control step, identified
by the authors as impossible to implement in practice. Deep reinforcement
learning addressed this intractability by replacing per-step DNS with
learned policies trained on custom reward signals, delivering
state-of-the-art drag reduction under realistic partial-state
sensing~\citep{guastoni_deep_2023}. Yet sustained online interaction
with high-fidelity DNS during training remains computationally
expensive~\citep{dulac-arnold_challenges_2019}, and the reliance on
hand-crafted reward proxies introduces a compounding limitation: proxy
objectives routinely diverge from the true control
goal~\citep{skalse_defining_2022, pan_effects_2021}, with prior remedies
through inverse reward design~\citep{hadfield-menell_inverse_2020} and
occupancy-measure regularisation~\citep{laidlaw_correlated_2024}
narrowing but not closing the gap.

\textbf{Generative models for physical systems. }
Score-based diffusion~\citep{song_score-based_2021} established generative modelling over continuous high-dimensional state spaces.
Subsequently, several applications confirmed that generative models can represent the complex statistics of turbulent PDE state spaces~\citep{lienen_zero_2024, vishwasrao_diff-sport_2025}.
Flow matching~\citep{lipman_flow_2023, albergo_building_2023, liu_flow_2023, tong_improving_2024} emerged as a simulation-free, ODE-based alternative, achieving comparable expressiveness at substantially lower inference cost through direct velocity field regression, with applications spanning protein backbones~\citep{bose_se3-stochastic_2024} and crystalline materials~\citep{miller_flowmm_2024}.
\citet{harder_efficient_2025} applied flow matching to turbulent PDE forecasting, comparing efficient sampling variants for autoregressive state prediction, but without any mechanism to steer generated states toward targeted reward configurations.

\textbf{Reward guidance in generative models. }
Classifier guidance \citep{dhariwal_diffusion_2021} and its classifier-free
variant \citep{ho_classier-free_nodate} established inference-time reward
steering as a tractable alternative to retraining, subsequently extended to
sequential decision-making through value-guided diffusion planners for offline
RL \citep{janner_planning_2022, ajay_is_2023} and reward alignment via direct
generative model fine-tuning \citep{black_training_2024, ren_diffusion_2024,
fan_online_2025}. In scientific design, \citet{uehara_reward-guided_2025} applied
inference-time guidance over diffusion to protein and DNA sequences,
and \citet{jensen_value_2026} proposed reward-guided flow matching via
online value-function learning for image generation and molecular design.
Both works generate standalone configurations in discrete or image domains.
The value-matching formulation of \citet{jensen_value_2026} cannot be
directly applied to turbulent velocity fields, as the required noise
injection destroys the spatial coherence of physical snapshots.
Inference-time guidance has not been applied over PDE-governed physical
state spaces or as targets for an active control policy.

\textbf{Learned surrogates and policy distillation. }
Learned dynamics models have replaced expensive simulators inside planning
loops across robotics \citep{nagabandi_neural_2017, williams_information_2017},
decoupling planning quality from per-step ground-truth access and enabling
tractable receding-horizon control in low-dimensional systems.
Policy distillation \citep{rusu_policy_2016} established the complementary
pattern of compressing an expensive planner into a fast reactive policy,
recently extended to offline behavioural supervision~\citep{qin_ipd_2026}.
Existing approaches rely on deterministic surrogate dynamics in low-dimensional
settings; no prior work has deployed either pattern in high-dimensional
turbulent flow with a generative planner capable of reward-steered inference.
\begin{figure}
    \centering
    \includegraphics[width=1\linewidth]{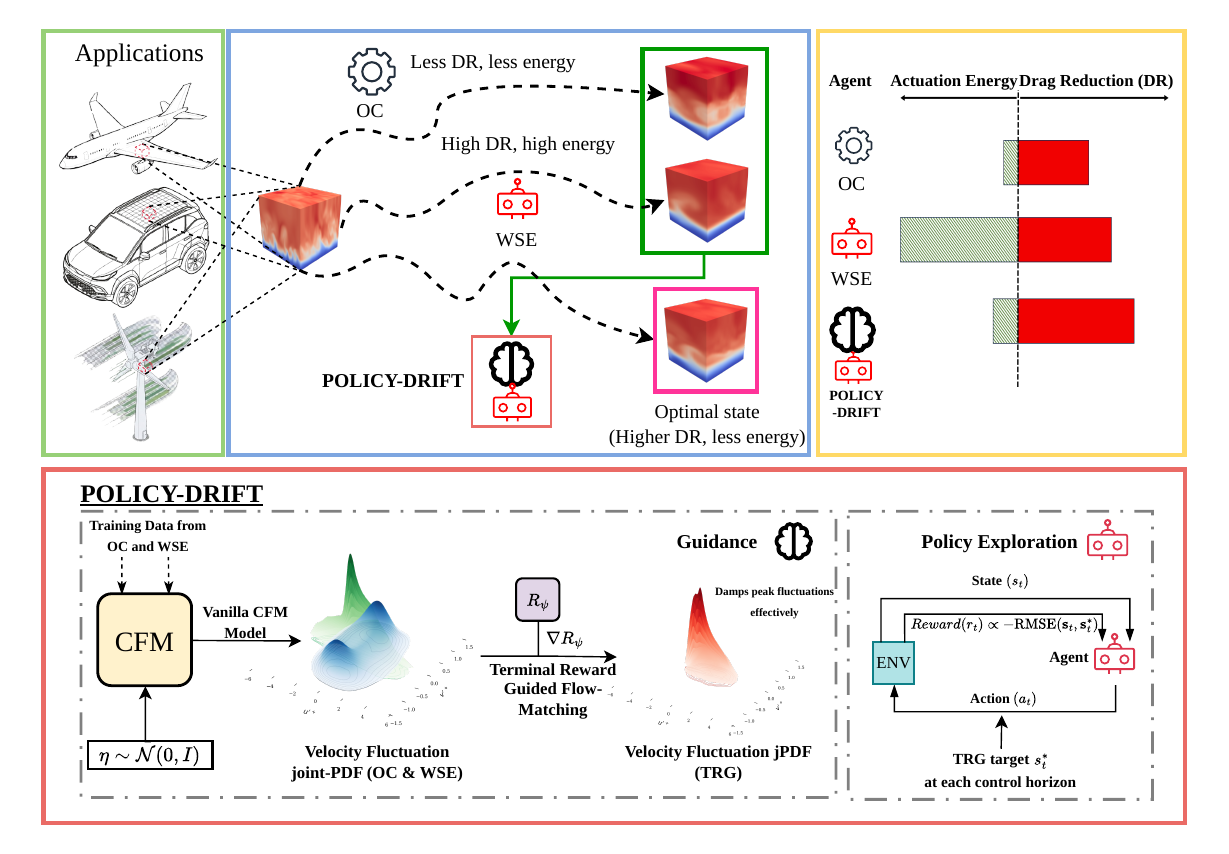}
     \caption{Policy-DRIFT framework. A CFM model trained on opposition control
         \citep{Choi_Moin_Kim_1994} and DRL-WSE \citep{guastoni_deep_2023}
         data is steered via Terminal Reward Guidance (TRG) to produce
         reward-maximising flow targets. A DRL agent tracks these targets
         via RMSE minimisation, achieving the highest drag reduction at
         optimised actuation energy (top right).}
  \label{fig:pipeline}
\end{figure}
\section{Methods}

\subsection{Problem Formulation}
\label{sec:problem}

We consider active drag reduction in turbulent open-channel flow of height $h$
at constant mass flux, driven by a time-varying pressure gradient.
Streamwise, wall-normal, and spanwise coordinates are $(x, y, z)$ with
corresponding velocity components $\mathbf{u} = (u, v, w)$.
The flow is governed by the incompressible Navier--Stokes equations,
\begin{align}
    \partial_t \mathbf{u} + (\mathbf{u} \cdot \nabla)\mathbf{u}
        &= -\nabla p + \frac{1}{\mathrm{Re_{cl}}} \Delta \mathbf{u}, \label{eq:NS} \\
    \nabla \cdot \mathbf{u} &= 0, \label{eq:continuity}
\end{align}
non-dimensionalised by $h$ and the laminar centreline velocity $U_{cl}$
($\mathrm{Re_{cl}} = U_{cl} h/\nu$).
No-slip and impermeability conditions are imposed at the lower wall, with
wall-normal actuation $v_w$ applied as a Dirichlet boundary condition.
Simulations use the spectral solver Dedalus~\citep{dedalus}, with Fourier
modes in $x$ and $z$ and Chebyshev polynomials in $y$, at resolution
$(N_x, N_z, N_y) = (16, 16, 64)$, with time integration via a third-order
DIRK+ERK scheme at $\delta t^+_{\rm sim} \approx 0.039$.
The friction Reynolds number is $\mathrm{Re}_\tau = u_\tau h / \nu = 180$,
where $u_\tau = \sqrt{\tau_w/\rho}$ is the friction velocity and $\nu$ the
kinematic viscosity, in a minimal channel
($L_x \times L_y \times L_z = 2.67h \times h \times 0.8h$).

The full flow state at any instant is
$\mathbf{u} \in \mathbb{R}^{3 \times N_x \times N_z \times N_y}$.
Long simulations ($t^+ > 20{,}000$) under three distinct control strategies
yield snapshot datasets $\{\mathcal{D}_k\}_{k=1}^{K}$, each consisting of
consecutive state pairs $(\mathbf{u}_0, \mathbf{u}_1) \in \mathcal{D}_k$
separated by a control horizon $\Delta t^+ = 5$ viscous time units
($t^+ = t/t^*$, $t^* = \nu/u_\tau^2$), with joint dataset
$\mathcal{D} = \bigcup_{k=1}^{K} \mathcal{D}_k$.
In this work $K = 3$: $\mathcal{D}_1$ (uncontrolled),
$\mathcal{D}_2$ (opposition-controlled, \citealt{Choi_Moin_Kim_1994}),
and $\mathcal{D}_3$ (wall-shear-stress DRL-controlled,
\citealt{guastoni_deep_2023}), with $|\mathcal{D}_k| = 7{,}350$ snapshot
pairs each.

\subsection{Multi-Regime Flow Manifold}
\label{sec:velocity-model}

We construct a generative model of the conditional distribution
$p_{\mathcal{D}}(\mathbf{u}_1 \mid \mathbf{u}_0)$ across all $K$ control regimes jointly.
For each target $\mathbf{u}_1$, define the conditional probability path
\begin{equation}
  p_s(\tilde{\mathbf{u}} \mid \mathbf{u}_1)
  = \mathcal{N}\!\left(\tilde{\mathbf{u}};\; s\,\mathbf{u}_1,\; (1-s)^2 I\right),
  \quad s \in [0,1],
  \label{eq:cond-path}
\end{equation}
which satisfies $p_0 = \mathcal{N}(\mathbf{0}, I)$ and $\lim_{s\to 1} p_s(\cdot \mid \mathbf{u}_1) = \delta_{\mathbf{u}_1}$.
A sample $\tilde{\mathbf{u}}_s \sim p_s(\cdot \mid \mathbf{u}_1)$ is realised as
$\tilde{\mathbf{u}}_s = (1-s)\boldsymbol{\eta} + s\,\mathbf{u}_1$ with $\boldsymbol{\eta} \sim \mathcal{N}(\mathbf{0}, I)$,
and the associated conditional vector field is
\begin{equation}
  u_s(\tilde{\mathbf{u}}_s \mid \mathbf{u}_1)
  = \frac{\mathbf{u}_1 - \tilde{\mathbf{u}}_s}{1 - s}
  = \mathbf{u}_1 - \boldsymbol{\eta},
  \label{eq:cond-vel}
\end{equation}
which is constant along each straight-line trajectory.
Conditioning further on the current state $\mathbf{u}_0$, the marginal probability path is
\begin{equation}
  p_s(\tilde{\mathbf{u}} \mid \mathbf{u}_0)
  = \int p_s(\tilde{\mathbf{u}} \mid \mathbf{u}_1)\, p_{\mathcal{D}}(\mathbf{u}_1 \mid \mathbf{u}_0)\,d\mathbf{u}_1,
  \label{eq:marginal-path}
\end{equation}
with marginal vector field
$u_s(\tilde{\mathbf{u}} \mid \mathbf{u}_0)
= \mathbb{E}\!\left[u_s(\tilde{\mathbf{u}}_s \mid \mathbf{u}_1) \mid \tilde{\mathbf{u}}_s = \tilde{\mathbf{u}},\, \mathbf{u}_0\right]$,
which is intractable because it requires integrating over the unknown
$p_{\mathcal{D}}(\mathbf{u}_1 \mid \mathbf{u}_0)$.
We parameterise $v_\theta(\tilde{\mathbf{u}}_s, s, \mathbf{u}_0) \approx u_s(\tilde{\mathbf{u}} \mid \mathbf{u}_0)$
and train it via the following result.

The conditional flow matching \citep{lipman_flow_2023, tong_improving_2024} objective
\begin{equation}
  \mathcal{L}_{\mathrm{CFM}}(\theta)
  = \mathbb{E}_{\substack{(\mathbf{u}_0,\mathbf{u}_1)\sim\mathcal{D} \\
                          \boldsymbol{\eta}\sim\mathcal{N}(\mathbf{0},I),\;
                          s\sim U[0,1]}}
  \!\left[\left\|v_\theta(\tilde{\mathbf{u}}_s,\, s,\, \mathbf{u}_0)
    - (\mathbf{u}_1 - \boldsymbol{\eta})\right\|^2\right]
  \label{eq:cfm-loss}
\end{equation}
and the marginal objective
$\mathcal{L}_{\mathrm{FM}}(\theta)
= \mathbb{E}_{(\mathbf{u}_0,\,s)}\,\mathbb{E}_{\tilde{\mathbf{u}}\sim p_s(\cdot\mid\mathbf{u}_0)}
\!\left[\|v_\theta(\tilde{\mathbf{u}}, s, \mathbf{u}_0) - u_s(\tilde{\mathbf{u}}\mid\mathbf{u}_0)\|^2\right]$
satisfy $\nabla_\theta\,\mathcal{L}_{\mathrm{CFM}} = \nabla_\theta\,\mathcal{L}_{\mathrm{FM}}$.

Minimising $\mathcal{L}_{\mathrm{CFM}}$ against the tractable conditional targets
$(\mathbf{u}_1 - \boldsymbol{\eta})$ therefore recovers $u_s(\cdot \mid \mathbf{u}_0)$
without computing the intractable integral in~\eqref{eq:marginal-path}.
Since $\mathcal{D} = \bigcup_{k=1}^{K}\mathcal{D}_k$, the expectation in~\eqref{eq:cfm-loss}
marginalises over all $K$ control regimes simultaneously;
no per-regime model, classifier head, or importance weight is required.
At inference, integrating
$d\tilde{\mathbf{u}}/ds = v_\theta(\tilde{\mathbf{u}}_s, s, \mathbf{u}_0)$
from $s=0$ to $s=1$ maps $\boldsymbol{\eta}\sim\mathcal{N}(\mathbf{0}, I)$
to a sample $\hat{\mathbf{u}}_1 \sim p_{\mathcal{D}}(\cdot \mid \mathbf{u}_0)$
drawn from the joint multi-regime distribution.

\subsection{Terminal Reward Guidance}
\label{sec:trg}

With $v_\theta$ frozen, the goal of TRG is to steer ODE trajectories toward terminal
states $\hat{\mathbf{u}}_1$ that maximise the scalar reward
\begin{equation}
  R(\hat{\mathbf{u}}_1)
  = \mathrm{DR}(\hat{\mathbf{u}}_1) - E_{\mathrm{act}}(\hat{\mathbf{u}}_1),
  \label{eq:reward}
\end{equation}
where
\begin{equation}
  \mathrm{DR} = 1 - \tau_w/\tau_{w,0}
  \label{eq:dr}
\end{equation}
is the drag-reduction fraction relative to uncontrolled flow
($\tau_w = \rho\nu\langle\partial_y u\rangle_{x,z}|_{\rm wall}$ is the mean wall-shear stress), and
\begin{equation}
  E_{\mathrm{act}} = \tfrac{1}{2}\langle|v_w|^3\rangle_{x,z} / u_\tau^3
  \label{eq:eact}
\end{equation}
is the normalised pumping energy of the wall actuation~\citep{kametani_effect_2015}.
This requires a model that can translate a gradient signal defined at the terminal
state back to intermediate ODE states, where it can influence the trajectory.

\paragraph{Reward Predictor.}
With $v_\theta$ fixed, generate a dataset of unguided ODE rollouts:
for each $(\mathbf{u}_0, \boldsymbol{\eta})$, integrate
$d\tilde{\mathbf{u}}/ds = v_\theta(\tilde{\mathbf{u}}_s, s, \mathbf{u}_0)$ to obtain
the trajectory $\{\tilde{\mathbf{u}}_{s_i}\}_{i=1}^{T}$ and terminal state
$\hat{\mathbf{u}}_1 = \tilde{\mathbf{u}}_1$, then compute the terminal reward
$r_T = R(\hat{\mathbf{u}}_1)$.
We train a predictor $R_\psi : \mathbb{R}^{3 \times N_x \times N_z \times N_y}
\times [0,1] \to \mathbb{R}$ to map any intermediate ODE state to $r_T$:
\begin{equation}
  \mathcal{L}_{\mathrm{RP}}(\psi)
  = \mathbb{E}_{i}\!\left[s_i \cdot \bigl(R_\psi(\tilde{\mathbf{u}}_{s_i}, s_i) - r_T\bigr)^2\right].
  \label{eq:rp-loss}
\end{equation}
The ramp weight $w(s) = s$ down-weights ODE states near $s=0$ (pure noise) and
assigns full weight near $s=1$ (physical field), where $R_\psi$ can be most
reliably trained.
Crucially, the same terminal $r_T$ serves as the regression target for every
intermediate state along the trajectory, with no one-step look-ahead
or Bellman bootstrap.
This eliminates compounding bootstrapping error at the cost of requiring $R_\psi$
to learn long-range reward prediction from early noisy states;
the ramp weighting mitigates this by concentrating supervision near the terminal end.

\paragraph{Guided ODE integration.}
At each ODE step from $s$ to $s + \delta s$, TRG applies a pre-placement
correction before the velocity-model(VM) step:
\begin{align}
  \tilde{\mathbf{u}}_s^{+} &= \tilde{\mathbf{u}}_s
    + \gamma\,\delta s\;\nabla_{\tilde{\mathbf{u}}_s} R_\psi(\tilde{\mathbf{u}}_s,\, s),
  \label{eq:pre-nudge} \\
  \tilde{\mathbf{u}}_{s+\delta s} &= \tilde{\mathbf{u}}_s^+
    + \delta s\cdot v_\theta\!\left(\tilde{\mathbf{u}}_s^{+},\, s,\, \mathbf{u}_0\right),
  \label{eq:vm-step}
\end{align}
where $\gamma > 0$ is the guidance scale.
The nudged intermediate state $\tilde{\mathbf{u}}_s^{+}$ is passed as input to $v_\theta$,
which propagates it through one learned velocity step. The alternative \emph{post-placement} scheme applies the correction after the VM step: $\tilde{\mathbf{u}}_{s+\delta s} = \tilde{\mathbf{u}}_s + \delta s\cdot v_\theta(\tilde{\mathbf{u}}_s, s, \mathbf{u}_0) + \gamma\,\delta s\;\nabla R_\psi$. In our experiments, post-placement causes reward hacking: drag reduction clamps at $\mathrm{DR} \to 1.0$ within the first few guidance steps, producing states that score high on $R_\psi$ but are physically unrealisable. Pre-placement avoids this because $v_\theta$ acts on $\tilde{\mathbf{u}}_s^{+}$ directly: the learned dynamics map the nudged intermediate back toward the support of $p_s(\cdot \mid \mathbf{u}_0)$, keeping the trajectory on the physical flow manifold at every step. This use of the generative model as an implicit manifold projector is the domain-specific design choice that distinguishes TRG from classifier guidance applied directly to the terminal state.

\subsection{Reinforcement-learning framework}
\label{sec:marl}

\textbf{Framework formulation }
We formulate closed-loop flow control as a reinforcement learning (RL) problem, in which several agents collectively optimise a shared global reward $r$ through distributed interaction with a fluid environment. A policy network is shared by agents, which maps partial flow observations $\mathbf{o}$ to continuous control actions $a$. This implementation reduces the effective parameter count and promotes generalisation through repeated exposure to diverse local flow states. We adopt TD3 \citep{fujimoto_td3_2018} as the policy algorithm for all
reward formulation comparisons; SAC \citep{haarnoja_sac_2018} is included as an additional off-policy
baseline under the WSE formulation, with algorithm selection justified in
Appendix~\ref{app:drl-training}.

The environment is partitioned into a set of pseudo-environments, each assigned to a single agent that observes and actuates its local subdomain independently. Here, the number of pseudo-environments is set equal to the number of active wall grid points $N_x \times N_z$. This yields a scalar action space $a \in \mathbb{R}^{1 \times 1}$ and an observation space of fixed dimension $\mathbf{o} \in \mathbb{R}^{N_{\rm feat} \times 1}$, where $N_{\rm feat}$ is the number of input features.

Each agent observes the inner-scaled fluctuation components of the streamwise ($u'^+$) and wall-normal ($v'^+$) velocity at an off-wall sensing plane $y^+ = 15$, giving $N_{\rm feat} = 2$. Prior to input, the spatial mean is subtracted from each feature over streamwise and spanwise directions in channel flow.
Actions correspond to wall-normal velocity Dirichlet boundary conditions, bounded by the local friction velocity $|a| \leq u_\tau$, modelling localised blowing and suction actuation. A zero-net-mass-flux (ZNMF) constraint is enforced by subtracting the spatial mean of the wall actuation before application, preventing spurious mass injection due to the incompressibility constraint.

\textbf{Reward formulations. }
We evaluate three reward functions. \textbf{WSE} uses $r = \mathrm{DR}$ (Eq.~\ref{eq:dr}), the instantaneous drag-reduction fraction relative to uncontrolled flow, following prior work~\citep{guastoni_deep_2023}. \textbf{WEN} extends this by penalising actuation energy: $r = \mathrm{DR} - E_{\mathrm{act}}$ (Eq.~\ref{eq:eact}). \textbf{TRG (Policy-DRIFT)} replaces the scalar reward entirely: the signal is the RMSE between the current flow state and the TRG-guided CFM target $\hat{\mathbf{u}}_1$, computed once at the start of each 8-step horizon ($\Delta t^+ = 5$) and held fixed throughout. Tracking a full-field spatial target also eliminates a pathological failure mode of scalar rewards: optimising instantaneous mean drag can exploit spatial averaging, yielding high rewards without genuine flow control, whereas RMSE to a full-field target enforces spatially uniform tracking without additional regularisation.

\section{Results}
\label{sec:results}
Policy-DRIFT is evaluated on turbulent channel flow DNS at $\mathrm{Re}_\tau = 180$ a canonical flow control configuration. Baselines include opposition control, the wall-shear-stress DRL controller (TD3-WSE) and a DRL controller trained directly on the cost-aware objective in Equation~\ref{eq:reward} (TD3-WEN), providing a direct comparison under identical reward specification. All methods are evaluated on six initial conditions held out from training. Results are reported sequentially: CFM fidelity, TRG guidance quality, and finally the closed-loop DRL performance.
\subsection{Conditional Flow Matching}
\label{sec:results-cfm}

A single conditional flow matching model $v_\theta$, trained without any regime labels or per-strategy supervision, learns to jointly represent the stochastic dynamics $p_\mathcal{D}(\mathbf{u}_1 \mid \mathbf{u}_0)$ across three qualitatively distinct control trajectories: passive uncontrolled evolution, opposition control, which selectively attenuates near-wall ejections and sweeps, and the aggressive wall-shear-stress DRL strategy that produces substantially different flow statistics. That these co-exist within a single learned manifold is not guaranteed: each strategy induces a different conditional distribution over successor states $\mathbf{u}_1$, and the model must resolve this multi-modal structure from the shared data $\mathcal{D}$. Quantitatively, $v_\theta$ achieves a mean squared error of order $10^{-4}$ on held-out snapshots from all three regimes, and the continuity residuals $|\nabla \cdot \hat{\mathbf{u}}_1|$ of the generated fields are of the same order of magnitude as those of the DNS snapshots, confirming that the joint training objective recovers the conditional vector field without per-regime supervision and that the generated states are physically realisable. The resulting manifold is richer than any individual strategy alone: by spanning the continuum between control regimes, it provides TRG with a space of physically realisable flow states that no single trajectory could supply.

\subsection{Terminal Reward Guidance (TRG)}
\label{sec:results-TRG}
\begin{figure}[ht]
  \centering
  \includegraphics[width=0.87\linewidth]{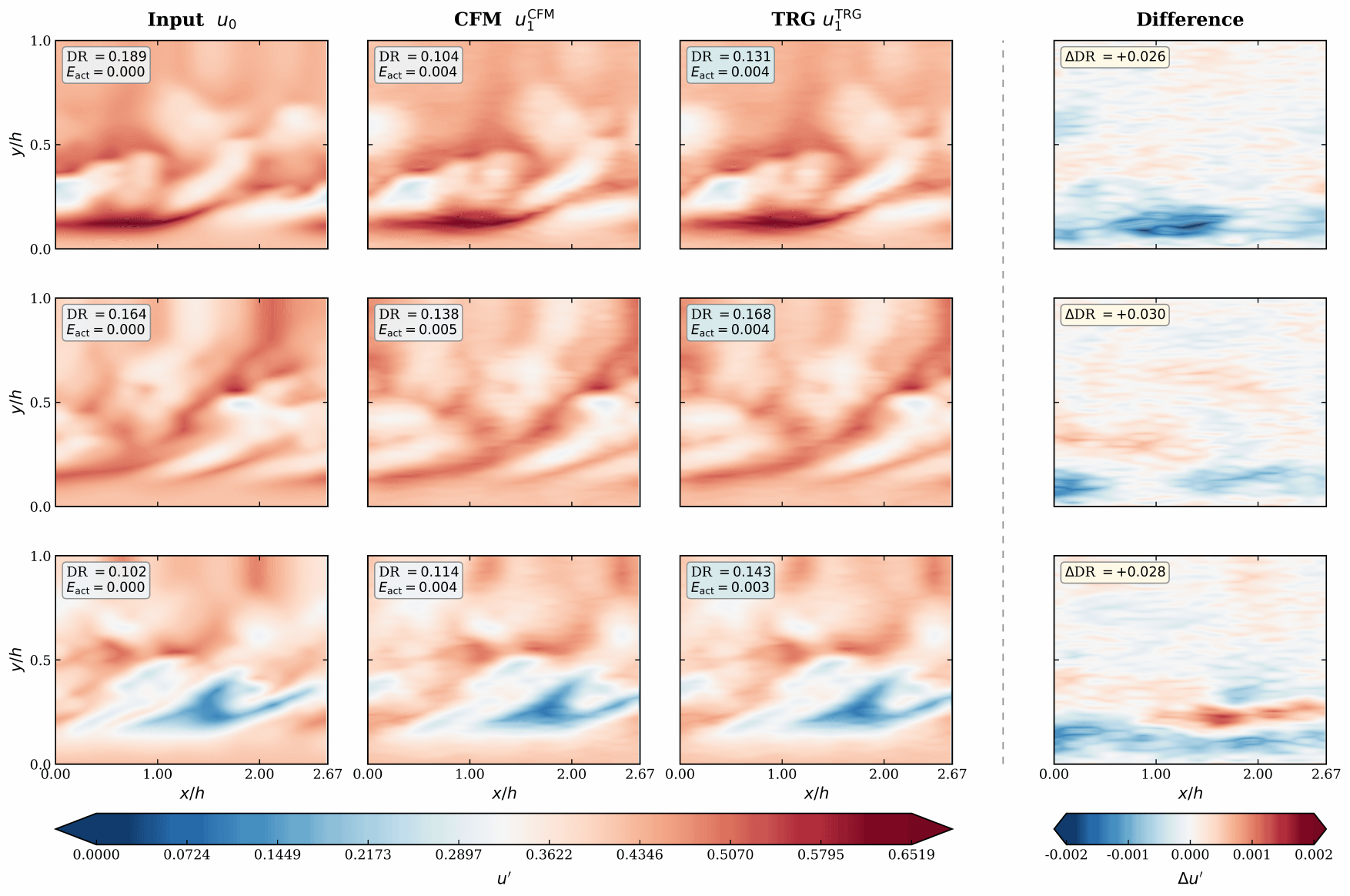}
  \caption{Streamwise velocity fluctuations $u'(x,y)$ in the $x$-$y$ plane for
three held-out uncontrolled ($\mathcal{D}_1$) snapshots. Columns show the
conditioning snapshot $\mathbf{u}_0$, the unguided CFM terminal state
$\mathbf{u}_1^{\rm CFM}$, the TRG-guided terminal state $\mathbf{u}_1^{\rm TRG}$,
and the absolute difference $\mathbf{u}_1^{\rm TRG} - \mathbf{u}_1^{\rm CFM}$.
TRG consistently improves drag reduction by $\Delta\mathrm{DR} \approx
0.026$--$0.030$ with negligible change in actuation energy. Corrections
concentrate near the wall ($y/h \approx 0.4$), where turbulent production
peaks, while the bulk flow remains largely undisturbed, the perturbation
magnitude is two orders smaller than the velocity fluctuations, confirming
physically-constrained guidance. Guidance scale $\gamma = 5$.}
\label{fig:guidance-uncontrolled}
\end{figure}

With $v_\theta$ fixed, TRG navigates the learned manifold toward its high-reward region at inference. Across all three held-out uncontrolled snapshots (Figure~\ref{fig:guidance-uncontrolled}), guided rollouts consistently reach higher drag reduction, by $\Delta\mathrm{DR} \approx 0.026$--$0.030$, relative to their unguided counterparts, while maintaining comparably low actuation energy, demonstrating that TRG extracts reward-maximising configurations already latent in the multi-regime manifold without introducing additional actuation cost. This behaviour is consistent across all three training regimes (Appendix~\ref{app:guidance-all-regimes}), confirming that the manifold spans reward-maximising states beyond what any individual control trajectory supplies.

The difference fields ($\mathbf{u}^{\rm TRG}_1 - \mathbf{u}^{\rm CFM}_1$) are informative about the mechanism: corrections concentrate near the wall ($y/h \approx 0.4$), precisely where turbulent production and skin-friction drag originate, while the bulk flow is largely undisturbed. The perturbation magnitude is two orders of magnitude smaller than the velocity fluctuations themselves, confirming that TRG acts as a targeted, physically-constrained nudge rather than a wholesale restructuring, a direct consequence of the pre-placement design, which routes guidance through $v_\theta$ at every step and keeps trajectories on the physical manifold. 

\subsection{Closed-Loop Control Performance}
\label{sec:results-drl}
\begin{figure}[h]
  \centering
  \includegraphics[width=0.85\linewidth]{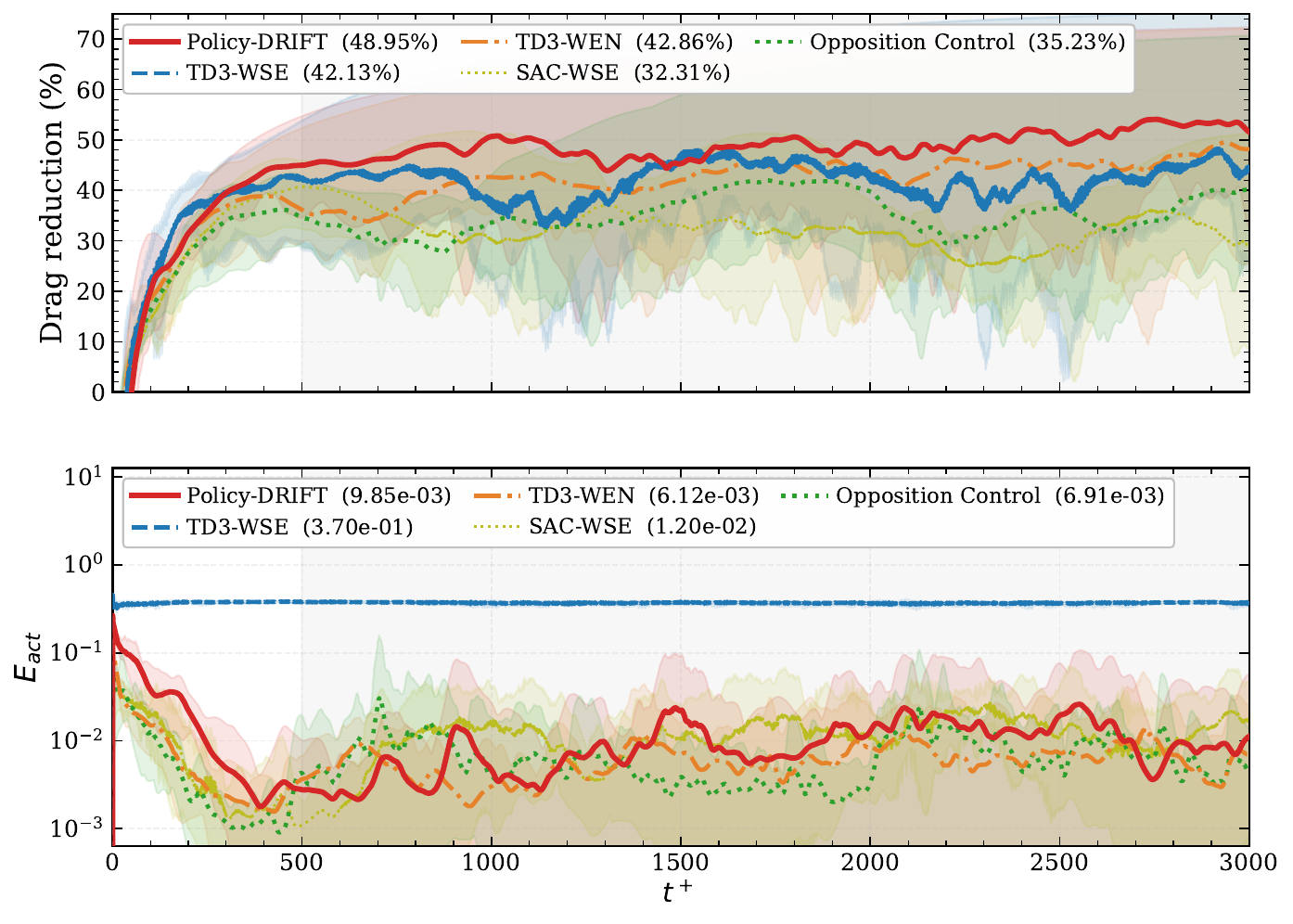}
  \caption{Drag reduction (top) and actuation energy $E_{\rm act}$ (bottom,
log scale) over $t^+ \in [0, 3000]$ for 6 out-of-distribution initial
conditions (shaded band: min--max across trajectories; scalar values
averaged over $t^+ \in [500, 3000]$). Policy-DRIFT separates from all
baselines by $t^+ \approx 200$ and sustains the highest drag reduction
with low variance throughout.}
  \label{fig:dr-energy-traj}
\end{figure}

Table~\ref{tab:drl-results} and Figure~\ref{fig:dr-energy-traj} summarise closed-loop performance across all methods. Policy-DRIFT achieves $48.95\%$ drag reduction, approaching the theoretical upper bound of $>50\%$ established by \citep{bewley2001dns} using full DNS access at every control step, while using only offline data and a single wall-parallel sensing plane at inference. Against the state-of-the-art DRL controller for flow control \citet{guastoni_deep_2023} (TD3-WSE), Policy-DRIFT gains $16.2\%$ in drag reduction while consuming ${\sim}37\times$ less actuation energy, improvements that emerge without the policy directly optimising either quantity. Compared to opposition control~\citep{Choi_Moin_Kim_1994}, the classical physics-based strategy that directly mitigates the near-wall momentum using full wall-normal velocity measurements, Policy-DRIFT achieves $38.9\%$ higher drag reduction at comparable actuation energy, demonstrating that a data-driven generative approach substantially outperforms classical heuristic flow control strategies. Figure~\ref{fig:dr-energy-traj} shows that this separation from baselines is established early in the transient ($t^+ \approx 200$) and maintained with low variance across all six out-of-distribution initial conditions, while TD3-WSE sustains actuation energy an order of magnitude above all other methods throughout.
\begin{table}[h]
  \centering
  \small
  \caption{Drag reduction (DR, \%) and normalised actuation energy
$E_{\mathrm{act}}$ averaged over 6 held-out initial conditions, with
standard deviation across trajectories. $\bar{R} = \mathrm{DR} -
E_{\mathrm{act}}$ is the cumulative cost-aware objective evaluated from
mean quantities. Training benchmarks (top) provide the data on which the
CFM is trained; all experimental variants (bottom) are evaluated under
identical conditions.}
  \label{tab:drl-results}
  \setlength{\tabcolsep}{4pt}
  \begin{tabular}{llccccc}
    \toprule
    \multirow{2}{*}{Category} & \multicolumn{1}{c}{\multirow{2}{*}{Method}} &
    \multicolumn{2}{c}{DR (\%)} &
    \multicolumn{2}{c}{$E_{\mathrm{act}}$ ($\times 10^{-1}$)} &
    \multirow{2}{*}{$\bar{R}$} \\
    \cmidrule(lr){3-4} \cmidrule(lr){5-6}
    & & Mean & Std & Mean & Std & \\
    \midrule
    \multirow{2}{*}{\shortstack[l]{Training\\Benchmarks}}
    & Opp-Control \citep{Choi_Moin_Kim_1994} & $35.23$ & $10.09$ & $0.0685$ & $0.0241$ & $0.345$ \\
    & TD3-WSE \citep{guastoni_deep_2023}     & $42.13$ & $12.71$ & $3.692$  & $0.0431$ & $0.052$ \\
    \midrule
    \multirow{4}{*}{\textit{Experiments}}
    & \textbf{Policy-DRIFT (Ours)} & $\mathbf{48.95}$ & $\mathbf{11.70}$ & $\mathbf{0.0990}$ & $\mathbf{0.0711}$ & $\mathbf{0.480}$ \\
    & TD3-WEN                      & $42.86$ & $11.77$ & $0.0609$ & $0.0406$ & $0.423$ \\
    & TD3-WSE                      & $42.13$ & $12.71$ & $3.692$  & $0.0431$ & $0.052$ \\
    & SAC-WSE                      & $32.31$ & $2.33$  & $0.1193$ & $0.0228$ & $0.311$ \\
    \bottomrule
  \end{tabular}
\end{table}

The most direct comparison is against TD3-WEN, which optimises the identical cost-aware objective $\mathrm{DR} - E_{\mathrm{act}}$ through policy gradients. Despite sharing the same reward, TD3-WEN achieves only $42.86\%$ DR, a $14.2\%$ gap closed by Policy-DRIFT for a modest $1.6\times$ increase in actuation energy. This gap directly isolates the cost of routing reward through policy gradients: the energy penalty enters every critic update, driving the policy toward suppressing actuation globally at the expense of drag reduction. In Policy-DRIFT, energy efficiency is never penalised by the policy; it emerges from the structure of the generative targets themselves. The reward acts where it is most effective, at target selection on the physical manifold, not at gradient computation.

\subsection{Physical Interpretation of Control Policies}

The joint PDF of $(u'^+, v'^+)$ at the sensing plane $y^+ = 15$, where turbulence production peaks~\citep{kim1987turbulence}, provides a physical picture of how each strategy acts on the near-wall cycle. The uncontrolled flow exhibits the characteristic broad distribution with pronounced ejection ($u'^+ < 0, v'^+ > 0$) and sweep ($u'^+ > 0, v'^+ < 0$) tails, the primary events responsible for skin-friction drag in wall-bounded turbulence. Opposition control narrows these tails substantially, confirming near-wall event suppression as its operating mechanism, but remains limited by its reliance on instantaneous wall-normal measurements without any learned flow representation. TD3-WSE pushes ejection and sweep suppression further but at the cost of a strongly bimodal $v'^+$ distribution extending to $\pm 1$, a clear over-actuation signature in which wall-normal advection amplifies rather than selectively attenuating turbulent events. TD3-WEN recovers a compact $v'^+$ distribution by penalising actuation energy, but the energy penalty suppresses actuation globally, weakening ejection and sweep attenuation relative to WSE. Policy-DRIFT resolves both failure modes simultaneously: the $v'^+$ footprint is as narrow as TD3-WEN while ejection and sweep suppression matches or exceeds TD3-WSE, combining strong drag reduction with energy efficiency without directly optimising either.

\begin{figure}[h]
  \centering
  \includegraphics[width=0.87\linewidth]{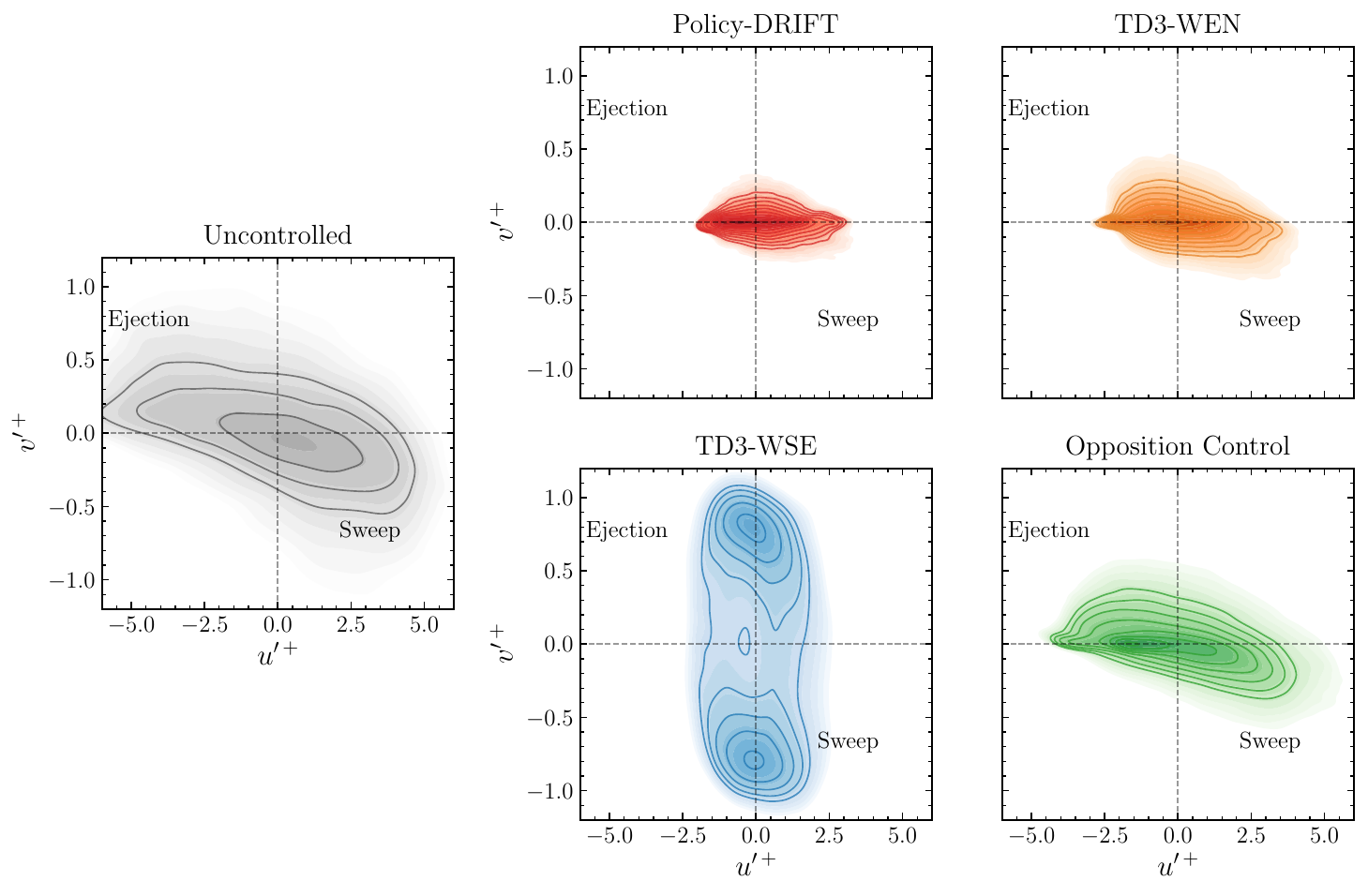}
  \caption{Joint PDFs of $(u'^+, v'^+)$ at $y^+ \approx 15$ for the
uncontrolled flow and all controlled methods. Dashed lines mark the
ejection ($u'^+ < 0,\, v'^+ > 0$) and sweep ($u'^+ > 0,\, v'^+ < 0$)
quadrants. Policy-DRIFT achieves the most compact distribution
with the strongest attenuation of both quadrants.}
  \label{fig:jpdf-y-15}
\end{figure}

\section{Conclusions}
\label{sec:Conclusions}
Policy-DRIFT demonstrates that relocating reward information from policy gradients
to generative inference systematically breaks the performance ceiling imposed by
reward misspecification. By tracking physically-grounded full-field targets rather
than scalar proxies, the policy avoids the structural failure modes of reward-based
DRL such as over-actuation, uninformative exploration, and surrogate-bounded performance
without any direct optimisation of the quantities it improves.

On turbulent channel flow at $\mathrm{Re}_\tau = 180$, Policy-DRIFT approaches the
theoretical upper bound of \citet{bewley2001dns} at $48.95\%$ drag
reduction, consuming ${\sim}37\times$ less actuation energy than the DRL baseline.
The CFM model requires no retraining when the control objective changes,
only $R_\psi$ need be updated, and the framework extends directly to
complex flow configurations such as aircraft wings \citep{wang2026physics},
opening a zero-shot pathway to drag reduction in geometries beyond the
training distribution.

\textbf{Limitations and future work.}
The present study is validated on a channel at $\mathrm{Re}_\tau = 180$; natural extensions include higher Reynolds numbers, complex wall geometries, and application to experimental data, a step toward digital-twin integration in fluid dynamics. Online adaptation of $R_\psi$ during closed-loop control and deployment across other high-fidelity physical domains where the same reward-structure bottleneck applies, such as plasma stabilisation, molecular dynamics, and robotics, are equally promising directions.

\section*{Acknowledgements}
This research was sponsored by the
Army Research Office and was accomplished under Grant Number W911NF-26-1-A053. The
views and conclusions contained in this document are those of the authors and should not be
interpreted as representing the official policies, either expressed or implied, of the Army
Research Office or the U.S. Government. The U.S. Government is authorized to reproduce and
distribute reprints for Government purposes notwithstanding any copyright notation herein.
\bibliographystyle{abbrvnat}
\bibliography{references}
\newpage
\appendix

\section{Data and Normalisation}
\label{app:data}

\paragraph{Dataset.}
The training data consists of three subsets of consecutive DNS snapshot
pairs $(\mathbf{u}_0, \mathbf{u}_1)$ at horizon $\Delta t^+ = 5$:
$\mathcal{D}_1$ (uncontrolled flow), $\mathcal{D}_2$ (opposition
control), and $\mathcal{D}_3$ (DRL wall-shear-stress control).
Each subset contains 7,350 pairs, yielding 22,050 training pairs in
the joint dataset $\mathcal{D}$.
An 80/20 train--test split is applied independently within each regime;
consecutive pairs never cross source boundaries.

\paragraph{Preprocessing.}
Raw DNS snapshots contain the instantaneous velocity field
$\mathbf{u} = (u, v, w) \in \mathbb{R}^{3 \times 16 \times 16 \times 64}$
(dimensions: components, $N_x$, $N_z$, $N_y$, with $N_y = 64$ wall-normal
points).
Velocity fluctuations are obtained by subtracting the per-voxel
time-mean $\bar{U}(x,y,z)$ estimated from the full dataset:
$\mathbf{u}' = \mathbf{u} - \bar{U}$.
Fluctuations are then normalised channel-wise with min-max scaling to
$[0,1]$ using statistics computed over the training set.
Table~\ref{tab:norm-stats} reports the per-channel statistics.

\begin{table}[h]
  \centering
  \caption{Per-channel min-max statistics used for normalisation
           ($\Delta t^+ = 5$ snapshot pairs).}
  \label{tab:norm-stats}
  \begin{tabular}{lccc}
    \toprule
    Channel & $\sigma$ & $\min$ & $\max$ \\
    \midrule
    $u'$ & 0.0548 & $-0.3726$ & $0.5168$ \\
    $v'$ & 0.0196 & $-0.2182$ & $0.2335$ \\
    $w'$ & 0.0261 & $-0.2723$ & $0.2987$ \\
    \bottomrule
  \end{tabular}
\end{table}

\section{CFM Velocity Model Architecture}
\label{app:arch-cfm}

The CFM velocity model $v_\theta$ is a 3D UNet operating on an input
volume of shape $[B, 7, 16, 16, 64]$, formed by concatenating the
linearly interpolated state $\tilde{\mathbf{u}}_s = (1-s)\boldsymbol{\eta} +
s\,\mathbf{u}_1$ (3 channels), the conditioning snapshot $\mathbf{u}_0$
(3 channels), and the scalar flow-matching time $s$ broadcast to the
spatial volume (1 channel).
The output is the predicted velocity field
$v_\theta \in \mathbb{R}^{[B,3,16,16,64]}$.

\paragraph{Architecture.}
The encoder has four levels with base width 64 and doubling channels
($64\!\to\!128\!\to\!256\!\to\!512$ at the bottleneck).
Each level applies two $3^3$ convolution blocks
(Conv3d $\to$ BatchNorm3d $\to$ ELU); spatial downsampling after each
of the first three levels uses $2^3$ average pooling.
The decoder mirrors this structure: each level upsamples
transposed-convolution, concatenates the corresponding encoder skip
connection, and applies a single convolution block.
The output layer uses a linear (identity) activation.
The total parameter count is 23.4M.

\begin{table}[h]
  \centering
  \caption{CFM UNet encoder--decoder channel widths and spatial
           resolutions. Input spatial shape: $16 \times 16 \times 64$.}
  \label{tab:unet-arch}
  \begin{tabular}{clcc}
    \toprule
    Level & Role & Channels & Spatial \\
    \midrule
    0 & Encoder (2 conv blocks) & $7    \to 64$  & $16 \times 16 \times 64$ \\
    1 & Encoder (2 conv blocks) & $64   \to 128$ & $8  \times 8  \times 32$ \\
    2 & Encoder (2 conv blocks) & $128  \to 256$ & $4  \times 4  \times 16$ \\
    3 & Bottleneck (2 conv blocks) & $256 \to 512$ & $2 \times 2 \times 8$  \\
    4 & Decoder                 & $512+256 \to 256$ & $4 \times 4 \times 16$ \\
    5 & Decoder                 & $256+128 \to 128$ & $8 \times 8 \times 32$ \\
    6 & Decoder                 & $128+64  \to 64$  & $16\times 16\times 64$ \\
    7 & Output                  & $64  \to 3$       & $16\times 16\times 64$ \\
    \bottomrule
  \end{tabular}
\end{table}

\paragraph{Training objective.}
The model minimises the conditional flow matching loss
\begin{equation}
  \mathcal{L}_{\mathrm{CFM}}
    = \mathbb{E}_{t,\boldsymbol{\eta},\mathbf{u}_1}
      \bigl\|v_\theta(\tilde{\mathbf{u}}_s, \mathbf{u}_0, t)
             - (\mathbf{u}_1 - \boldsymbol{\eta})\bigr\|^2,
\end{equation}
where $\boldsymbol{\eta}\sim\mathcal{N}(0,I)$,
$t\sim\mathrm{Uniform}(0,1)$, and the target velocity
($\mathbf{u}_1 - \boldsymbol{\eta}$) is constant along each
straight-line OT path.

\paragraph{Training details.}
Optimiser: AdamW with learning rate $10^{-3}$, weight decay $10^{-2}$,
gradient clipping at 1.0, and a ReduceLROnPlateau scheduler (patience
20, factor 0.5).
Training runs for up to 500 epochs with early stopping (patience 50).
Batch size is 256.
ODE integration at inference uses 50 Euler steps.

\section{Reward Predictor Architecture}
\label{app:arch-rp}

The reward predictor $R_\psi(x_s, s)$ is a 3D CNN with four
convolutional stages and Feature-wise Linear Modulation (FiLM) time
conditioning.

\paragraph{Time embedding.}
The scalar ODE time $s \in [0, 1]$ is encoded as a 64-dimensional
sinusoidal embedding:
\begin{equation}
  \mathrm{emb}(s)
  = \bigl[\sin(s\,f_k),\;\cos(s\,f_k)\bigr]_{k=0}^{31},
  \quad f_k = 10{,}000^{-k/32},
\end{equation}

\paragraph{FiLM conditioning.}
At each stage the embedding is linearly projected to a scale--shift
pair $(\gamma, \beta) \in \mathbb{R}^{2}$ and applied after group
normalisation: $h \leftarrow \gamma \odot \mathrm{GroupNorm}(h) + \beta$.

\paragraph{Network stages.}
Four convolutional stages with base width $h = 64$ (1.29M parameters total):

\begin{table}[h]
  \centering
  \caption{Reward predictor stage-wise architecture.
           Input shape: $[B, 3, 16, 16, 64]$.}
  \label{tab:rp-arch}
  \begin{tabular}{clcccc}
    \toprule
    Stage & Kernel & Stride & Ch.\ in $\to$ out & Output spatial & Norm \\
    \midrule
    0 & $3^3$ & 1 & $3   \to 64$  & $16 \times 16 \times 64$ & GN(8) \\
    1 & $3^3$ & 2 & $64  \to 64$  & $8  \times 8  \times 32$ & GN(8) \\
    2 & $3^3$ & 1 & $64  \to 128$ & $8  \times 8  \times 32$ & GN(8) \\
    3 & $3^3$ & 1 & $128 \to 256$ & $8  \times 8  \times 32$ & GN(8) \\
    \midrule
    Head & $1^3$ & -- & $256 \to 1$ & $8 \times 8 \times 32$ & -- \\
    \bottomrule
  \end{tabular}
\end{table}

After stage 3, a $1^3$ convolution produces a spatial score map of
shape $[B, 1, 8, 8, 32]$; summing its $8\!\times\!8\!\times\!32 = 2{,}048$
values yields the scalar prediction $R_\psi \in \mathbb{R}^B$.
The head weights are initialised by dividing by 2{,}048.
All stages use SiLU activations.

\paragraph{Training.}
$R_\psi$ is trained with AdamW (learning rate $10^{-4}$, weight decay
$10^{-2}$, gradient clipping at 1.0) and a cosine annealing schedule
over 500 epochs with early stopping (patience 50), batch size 128.
Unguided ODE rollouts from the frozen $v_\theta$ are generated using
50 Euler steps split into $T = 25$ segments; the terminal reward $r_T$
is assigned as the regression target to each intermediate state
$\tilde{\mathbf{u}}_{s_i}$, minimising
\begin{equation}
  \mathcal{L}_{\mathrm{RP}}(\psi)
  = \mathbb{E}_{i}\!\left[s_i \cdot
    \bigl(R_\psi(\tilde{\mathbf{u}}_{s_i}, s_i) - r_T\bigr)^2\right],
\end{equation}
where the linear ramp $w(s_i) = s_i$ concentrates gradient signal near
$s = 1$, where ODE states closely resemble physical velocity snapshots.
At inference, the ODE is steered via $\nabla_{\tilde{\mathbf{u}}} R_\psi$
using the same 50 Euler steps.

\paragraph{Inference.}
At each ODE step, the gradient $\nabla_{\tilde{\mathbf{u}}} R_\psi(\tilde{\mathbf{u}}, s)$ steers the
frozen integrator toward high-reward terminal states.
With \emph{pre-guidance} (default), the correction is applied before
the flow model:
\begin{equation}
  \tilde{\mathbf{u}}_s^{+} = \tilde{\mathbf{u}}_s + \gamma\,\delta s\;\nabla_{\tilde{\mathbf{u}}_s} R_\psi(\tilde{\mathbf{u}}_s, s),
  \qquad
  \tilde{\mathbf{u}}_{s+\delta s} \leftarrow \tilde{\mathbf{u}}_s^+ + v_\theta(\tilde{\mathbf{u}}_s^{+}, \mathbf{u}_0, s)\,\delta s,
\end{equation}
so the flow model acts as a regulariser on the perturbation.
Setting $\gamma = 0$ recovers the unguided base ODE.
The guidance scale $\gamma$ is a free inference hyperparameter
unconstrained by any noise
schedule.

\section{TRG Guidance on All Regimes}
\label{app:guidance-all-regimes}

\begin{figure}[h]
  \centering
  \includegraphics[width=0.9\linewidth]{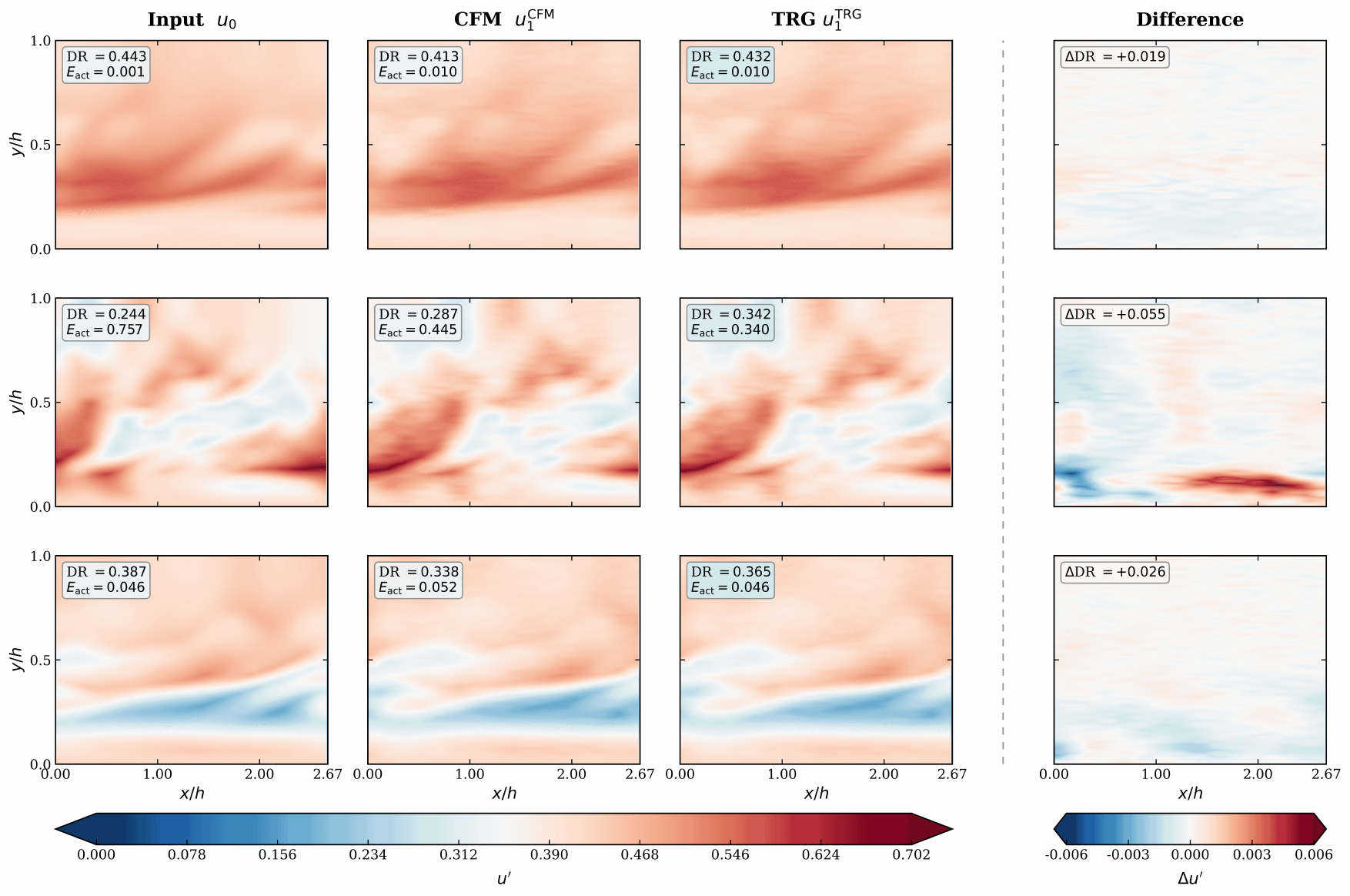}
  \caption{Streamwise velocity fluctuations $u'(x,y)$ in the $x$-$y$ plane for three
held-out opposition-controlled ($\mathcal{D}_2$) snapshots. Columns show the
conditioning snapshot $\mathbf{u}_0$, the unguided CFM terminal state
$\mathbf{u}_1^{\mathrm{CFM}}$, the TRG-guided terminal state
$\mathbf{u}_1^{\mathrm{TRG}}$, and the absolute difference
$\mathbf{u}_1^{\mathrm{TRG}} - \mathbf{u}_1^{\mathrm{CFM}}$, with reward
values (DR, $E_{\mathrm{act}}$) annotated per panel. Guidance scale
$\gamma = 5$.}
  \label{fig:guidance-OC}
\end{figure}

\begin{figure}[h]
  \centering
  \includegraphics[width=0.9\linewidth]{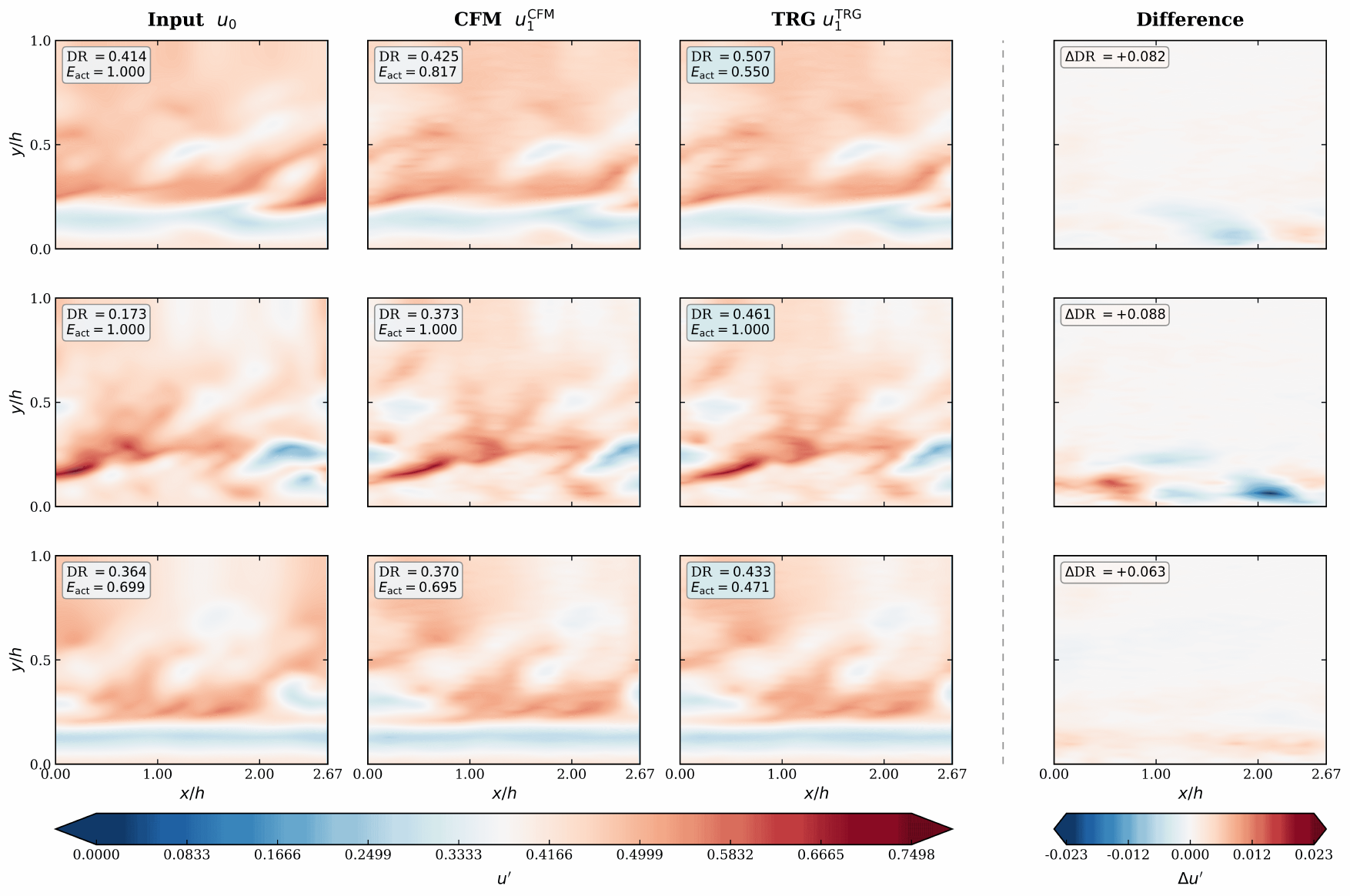}
  \caption{Streamwise velocity fluctuations $u'(x,y)$ in the $x$-$y$ plane for three
held-out TD3-WSE ($\mathcal{D}_3$) snapshots. Columns show the conditioning
snapshot $\mathbf{u}_0$, the unguided CFM terminal state
$\mathbf{u}_1^{\mathrm{CFM}}$, the TRG-guided terminal state
$\mathbf{u}_1^{\mathrm{TRG}}$, and the absolute difference
$\mathbf{u}_1^{\mathrm{TRG}} - \mathbf{u}_1^{\mathrm{CFM}}$, with reward
values (DR, $E_{\mathrm{act}}$) annotated per panel. The energy penalty
is more pronounced than in $\mathcal{D}_2$, reflecting the higher baseline
actuation of the DRL-WSE regime. Guidance scale $\gamma = 5$.}
  \label{fig:guidance-DRL-WSE}
\end{figure}

Figures~\ref{fig:guidance-OC} and~\ref{fig:guidance-DRL-WSE} extend Figure~\ref{fig:guidance-uncontrolled}
to opposition-controlled ($\mathcal{D}_2$) and DRL-WSE ($\mathcal{D}_3$)
initial conditions.
In all regimes, guided rollouts achieve higher terminal reward than
unguided rollouts, with the improvement consistent across independent
noise initialisations. The energy penalty is fairly noticeable for opposition-controlled and DRL-WSE with guided ODE rollouts resulting in a significant reduction in actuation energy.

\newpage
\section{DRL Training Details}
\label{app:drl-training}
\paragraph{Training configuration}
Training is conducted across six distinct initial conditions over 100 episodes. At the start of each episode, an initial condition is selected via a seeded random number generator to ensure reproducibility. Each episode spans a simulation time of $t^+ = 1{,}500$, scaled by the viscous time unit ($t^* = \nu / u_\tau^2$, $t^+ = t/t^*$), and the policy is updated every $\Delta t^+_{\rm upd} = 180$ using 64 mini-batch gradient steps. The control policy is evaluated at intervals of $\delta t^+ = 0.6$, yielding 2{,}500 actuation updates per episode. Mini-batches of size 64 are sampled from a replay buffer of capacity $10^6$ containing state-action-reward tuples. Exploration is promoted by adding zero-mean Gaussian noise with standard deviation $0.2\,u_\tau$ to the actor output prior to critic evaluation. All remaining hyperparameters are set to the default values of the Stable-Baselines3 implementation~\citep{stable-baselines3}.

\paragraph{Policy algorithm}
We adopt the model-free, off-policy twin-delayed deep deterministic policy gradient (TD3)~\citep{fujimoto_td3_2018} to train the DRL controller. TD3 mitigates the Q-value overestimation bias of DDPG~\citep{lillicrap_ddpg_2015} through clipped double Q-learning, and has demonstrated effectiveness in recent turbulent flow-control studies~\citep{lee_turbulence_2023,sonoda_tcfdrl_2023,guastoni_deep_2023,walchli_drag_2024,cavallazzi_deep_2025,zhou_reinforcement-learning-based_2025}. The off-policy design is particularly suited to wall-bounded turbulence:
TD3 accumulates transitions in a replay buffer and reuses them across
multiple updates, providing a statistically richer coverage of the chaotic
flow dynamics without simulation cost. For comparison, we also report
results for soft actor-critic (SAC)~\citep{haarnoja_sac_2018} under the
WSE reward formulation. SAC exhibited high episode variance and divergent
behaviour in later training phases, consistent with known instabilities
of entropy-regularised methods under sparse reward signals, as reflected
in the lower drag reduction reported in Table~\ref{tab:drl-results};
TD3 is therefore used for all reward formulation comparisons. The configurations are summarized in Table~\ref{tab:drl-results}.

Table~\ref{tab:drl-training} summarises all hyperparameters used for DRL training and evaluation. Note that we applied the default parameters by StableBaselines3~\citep{stable-baselines3} for the rest.

\begin{table}[h]
\centering
\caption{Policy architecture and training hyperparameters.}
\label{tab:drl-training}
\begin{tabular}{llll}
\toprule
Category & Parameter & Value & Notes \\
\midrule
\multirow{2}{*}{Architecture} & Critic hidden widths & $[16,\,64,\,64]$ & MLP, ReLU \\
                              & Actor hidden width   & $[8]$            & MLP, ReLU \\
\midrule
\multirow{2}{*}{Common}       & Training episodes    & $400$            & — \\
                              & Learning rate        & $10^{-4}$        & — \\
\midrule
\multirow{3}{*}{Off-policy}   & Replay buffer size   & $10^6$           & TD3, SAC \\
\multirow{3}{*}{(TD3/SAC)}    & Batch size           & $64$             & — \\
                              & Gradient updates     & $64$             & Every 300 env.\ steps \\
\midrule
Checkpoint & Criterion & Mean episode reward & Held-out initial condition \\
\bottomrule
\end{tabular}
\end{table}

\section{Guidance Scale Ablation}
\label{app:ablation-gamma}
We sweep the TRG guidance scale $\gamma \in \{5, 10, 20, 30\}$ and train
separate DRL policies for each value, evaluating mean DR\% and
$E_{\mathrm{act}}$ over six held-out initial conditions.

As $\gamma$ increases, the guidance signal exerts stronger influence over the
ODE trajectory, progressively shifting the terminal states toward lower
actuation energy at the cost of drag reduction. However, the energy savings
do not compensate for the loss in drag reduction: moving from $\gamma = 5$ to
$\gamma = 20$ reduces $E_{\mathrm{act}}$ by a factor of approximately $2.4$
while sacrificing $12.4$ percentage points of drag reduction. We therefore
select $\gamma = 5$ as it achieves the highest drag reduction while maintaining
low actuation energy. Table~\ref{tab:gs-sweep} summarises the results.

\begin{table}[h]
  \centering
  \caption{Guidance scale sweep for Policy-DRIFT: drag reduction (DR, \%) and
           mean actuation energy $E_{\mathrm{act}}$ averaged over 6
           held-out initial conditions. The selected value is highlighted.}
  \label{tab:gs-sweep}
  \begin{tabular}{lcccc}
    \toprule
    \multirow{2}{*}{Guidance Scale ($\gamma$)} &
    \multicolumn{2}{c}{DR (\%)} &
    \multicolumn{2}{c}{$E_{\mathrm{act}}$ ($\times 10^{-1}$)} \\
    \cmidrule(lr){2-3} \cmidrule(lr){4-5}
    & Mean & Std & Mean & Std \\
    \midrule
    \textbf{5}  & $\mathbf{48.95}$ & $\mathbf{11.70}$ & $\mathbf{0.0990}$ & $\mathbf{0.0711}$ \\
    10          & $47.22$          & $10.61$          & $0.1549$          & $0.1104$          \\
    20          & $36.58$          & $7.18$           & $0.0419$          & $0.0140$          \\
    30          & $35.66$          & $6.78$           & $0.0533$          & $0.0165$          \\
    \bottomrule
  \end{tabular}
\end{table}

\newpage
\section{Computational Resources}
\label{appendix:compute}

All experiments were conducted on an HPC cluster. Two node types were employed:

\paragraph{GPU node (CFM training)} One node was used for all CFM training runs: dual-socket x86\_64 architecture, 64 physical cores (2$\times$32-core sockets, 1 thread per core), 1{,}512\,GB RAM, and 4$\times$ NVIDIA H200 GPUs. 

\paragraph{CPU node (CFD simulation and DRL training)} One node was used for flow simulations: dual-socket x86\_64 architecture, 192 physical cores (2$\times$96-core sockets, 1 thread per core), 373\,GB RAM.

Each experimental run used a single node. DRL training was performed on the GPU node, while turbulent flow simulations, DRL training and inference were executed on the CPU node.
The conditional flow matching (CFM) model
converged in approximately \textbf{2.5 hours} (500 epochs) on a single NVIDIA H200 GPU.
The reward predictor (TRG) converged in approximately \textbf{10 hours} (100 epochs)
using a single NVIDIA H200 GPU.   

\section{Licenses}
The entire codebase will be open-sourced upon publication in compliance 
with the licenses of all packages used. The relevant licenses are listed below:
\begin{enumerate}
    \item \textbf{Dedalus} — GNU GPL v3.0
    \item \textbf{PyTorch} — BSD 3-Clause License
    \item \textbf{Stable-Baselines3} — MIT License
\end{enumerate}

\end{document}